\documentclass[12pt]{article}
\usepackage{graphicx}
\usepackage{amsfonts}
\usepackage{amssymb}
\usepackage{mathrsfs}
\usepackage{amsmath}
\usepackage[unicode]{hyperref}
\usepackage[numbers,sort&compress]{natbib}

\begin{document}
\renewcommand{\thefootnote}{\fnsymbol{footnote}}
\begin{titlepage}

\vspace{10mm}
\begin{center}
{\Large\bf Spacetime quantization effects on 5-dimensional black string evaporation}
\vspace{8mm}

{{\large Xiang-Qian Li${}^{1}$\footnote{E-mail address: lixiangqian13b@mails.ucas.ac.cn}}\\

\vspace{6mm}
${}^{1}${\normalsize \em School of Physics, University of Chinese Academy of Sciences, Beijing 100049, China}

}
\end{center}

\vspace{10mm}
\centerline{{\bf{Abstract}}}
\vspace{6mm}
Spacetime quantization predicts the existence of minimal length and time-interval. Within 5-dimensional Schwarzschild-like black string background, the tunneling of scalar particles, fermions and massive bosons are first studied together in the same generalized uncertainty principle framework. It is found that, the minimal length and time-interval effect weakens the original Hawking radiation. To $\mathcal{O}(\frac{1}{M_f^2})$, the corrected temperatures depend on not only the mass of black string, but also the mass and angular momentum of emitted particles. The temperature correction for massive bosons is four times as big as that for scalar particles and fermions. As a result, the bosons cease to tunnel from the black string before the scalar particles and fermions do. The evaporation remnant is expected in our analysis, however it should be verified by full quantum gravity theory.
\vskip 20pt

\noindent

\end{titlepage}
\newpage
\renewcommand{\thefootnote}{\arabic{footnote}}
\setcounter{footnote}{0}
\setcounter{page}{2}

\section{Introduction}
\label{Introduction}
Hawking radiation is an important phenomenon within black hole physics. It states that black hole can release radiation, due to quantum effect near the event horizon~\cite{hawking}. Various methods have been developed to study Hawking radiation, among which two kinds of tunneling methods, null geodesic method and Hamilton-Jacobi method, are the most popular ones. Both of these two approaches to tunneling use the fact that the tunneling probability is related to the imaginary part of the classically forbidden trajectory from inside to outside the horizon by $\Gamma\propto {\rm exp}(-2{\rm Im}S/\hbar)$, with WKB approximation, that $S$ is the classical action of the trajectory to leading order in $\hbar$, applied. The null geodesic method was constructed by Parikh and Wilczek~\cite{mkpf,mkp1,mkp2}; The Hamilton-Jacobi method was developed by Angheben \textit{et al}~\cite{mamn} based on Padmanabhan's works~\cite{kstp,sskt}. Using these two methods, different kinds of particles tunneling from various backgrounds have been investigated~\cite{ecv,ajmm,maam,zzhao1,zzhao2,rkrbm,ajmme,pmit,dysz,kern1,kern2,Chen:2011mg,Kruglov:2014iya,sik,grch1,grch2,grch3,xqli,Sakalli1,Li:2016sgj,Sakalli3}.

One common feature among various quantum gravity theories, such as string theory, loop quantum gravity, and noncommutative geometry, is the existence of a minimum measurable length~\cite{kppplb,maggi,garayIJMP,Scardigli:1999jh,amelinoIJMP}. An effective model to realize the minimal length is the generalized uncertainty principle (GUP), based on which the first generalized uncertainty relation was proposed by~\cite{Kempfprd}. Considering both the minimal length effect and minimal time-interval effect, a model of momentum-wave vector and energy-frequency relations which incorporates the central idea of Large eXtra Dimensions (LXDs), is given by~\cite{Hossenfelder:2003jz}
\begin{eqnarray}
\label{modifiedp}p_i&\approx& p_{i0}\left(1+\beta p^2_{i0}\right),\\
\label{modifiedE}E&\approx& E_0\left(1+\beta E^2_0\right),
\end{eqnarray}
where $p_{i0}=-i\hbar\partial_i$ and $E_0=i\hbar\partial_t$ are operators within Heisenberg uncertainty principle (HUP) framework, $\beta=1/(3M_f^2)$ with $M_f$ representing the higher dimensional Planck mass. The current lower limits on $M_f$ range from 3.67 Tev/$c^2$ for 2 compactified LXDs and to 2.25 Tev/$c^2$ for 6 compactified LXDs~\cite{Chatrchyan:2012me}. The relations in Eqs.~\eqref{modifiedp} and~\eqref{modifiedE} exist as the low energy limit ($p\ll M_f$) approximation of the full $p_i(p_{0i})$ and $E(E_0)$ relations, thus are not Lorentz covariant. Then the modified commutation relation yields
\begin{equation}\label{commutation}
\left[x_i,p_j\right]= i\hbar \delta_{ij} \left(1+ \beta p^2_i\right),
\end{equation}
and the corresponding generalized uncertainty relation is
\begin{equation}\label{GUP}
\Delta x_{i} \Delta p_{j} \geq \frac{\hbar}{2}\delta_{ij}\left[1+ \beta \langle p^2_i\rangle\right].
\end{equation}

Generalized uncertainty principle is an effective tool to relate black hole with quantum gravity properties. A lot of papers study black hole physics with GUP incorporated. The thermodynamics of black holes has been investigated in the framework of GUP~\cite{aliJHEP,majumPLB,binaPRD,chenNPPS,adlerGRG,xiangwenJHEP,kimsonJHEP}. Combining the GUP with the null geodesic tunneling method, Nozari and Mehdipour studied the modified tunneling rate of the Schwarzschild black hole~\cite{nozariEPL}. The GUP deformed Hamilton-Jacobi equations for fermions and scalar particles in curved spacetime have been introduced and the corrected Hawking temperatures have been derived for various spacetime in~\cite{majumGRG,chenahep,chenjhep,chenjcap,barguenoPLB,mubenrong,pengwang,anaclePLB,fengEPJC}. Taking the GUP scheme, which incorporates the central idea of Large eXtra Dimensions, into account, the tunneling processes of massive bosons ($W^{\pm}$, $Z^0$) from Reissner-Nordstrom and Kerr black holes were investigated~\cite{Li:2016mwq}.

The LXDs model supposes the existence of extra compactified dimensions. In this paper, we investigate scalar particles, fermions and massive bosons tunneling across the horizons of 5-dimensional black string using the Hamilton-Jacobi method which incorporates the minimal length and time-interval (MLT) effect via Eqs.~\eqref{modifiedp} and~\eqref{modifiedE}. Including the additional dimension $\omega$ to the metric of the Schwarzschild spacetime, the obtained black string solution takes the following form
\begin{equation}\label{metric}
\text{d}s^2=-F(r)\text{d}t^2+F(r)^{-1}\text{d}r^2+r^2(\text{d}\theta^2+\text{sin}^2{\theta}\text{d}\phi^2)+\text{d}\omega^2,\quad F(r)=1-\frac{2M}{r}.
\end{equation}
Our calculation shows that the quantum gravity correction is related not only to the black string's mass but also to the masses and angular momentums of
emitted particles. The quantum gravity correction explicitly decelerates the temperature increasing in the
process of black string evaporation. What's more, the temperature correction for massive bosons is four times as big as that for scalar particles and fermions. As a natural result, the black string cease to emit massive bosons earlier than ceasing to emit scalar particles and fermions. We also discuss the black string thermodynamics and the remnant of evaporation, concluding that a full quantum gravity theory is needed to study the final stage of the black string evaporation. People may question the validity of applying WKB approximation to considering the quantum gravity effect, since the WKB approximation is a semi-classical approximation retaining only the leading order in $\hbar$ and the quantum gravity effect related to the Planck scale should be tiny in low energy limit. We show that, it is reasonable to apply WKB approximation in our analysis, considering the specific process with which the quantum gravity effect influences the black string evaporation.

The organization of this paper is as follows. Incorporating the MLT effect, the tunneling process of scalar particles, fermions and massive bosons are investigated in section~\ref{Section2},~\ref{Section3} and~\ref{Section4}, respectively. The thermodynamics of black string are discussed in section~\ref{Section5}. Section~\ref{Section6} is devoted to our discussion and conclusion. Appendices~\ref{appendix1} is added to justify the application of WKB-approximation to studying the effects of quantum gravity on particles tunneling process. We use the spacelike metric signature convention $(-,+,+,+)$; $G=c=k_B=1$ is set through this paper except in section~\ref{Section5}, where we use IU to make the physical meaning more explicit.

\section{Tunneling process of scalar particles}
\label{Section2}

Within the framework of generalized uncertainty principle, the modified Klein-Gordon equation in flat spacetime yields~\cite{Hossenfelder:2003jz}
\begin{equation}\label{gKGe1}
\hbar^2 \eta^{\mu\nu}\left(\partial_{\mu}-\beta {\hbar}^2\partial_{\mu}^3\right)\left(\partial_{\nu}-\beta {\hbar}^2\partial_{\nu}^3\right)\phi=m^2\phi.
\end{equation}
Generalizing it to curved spacetime with diagonal metric components, one obtains
\begin{equation}\label{gKGe2}
\hbar^2 g^{\mu\nu}\left[\partial_{\mu}+(-1)^{1+\delta^0_{\mu}}\beta {\hbar}^2g^{\mu\mu}\partial_{\mu}^3\right]\left[\partial_{\nu}+(-1)^{1+\delta^0_{\nu}}\beta {\hbar}^2g^{\nu\nu}\partial_{\nu}^3\right]\phi=m^2\phi,
\end{equation}
where the difference of signs of the $\mathcal{O(\beta)}$ term for time component and that for space components in square brackets is originated from the fact that $g^{00}$ always shares different sign with $g^{ii}$.

Applying the WKB approximation by assuming an ansatz of the form of wave function
\begin{equation} \label{WKB1}
\phi={\rm exp}\left[\frac{i}{\hbar}S(t,r,\theta,\phi,\omega)\right],
\end{equation}
where $S$ is defined as
\begin{equation} \label{S0123}
S(t,r,\theta,\phi,\omega)=S_0(t,r,\theta,\phi,\omega)+\hbar S_1(t,r,\theta,\phi,\omega)+\hbar^2 S_2(t,r,\theta,\phi,\omega)+\cdots.
\end{equation}

Inserting Eqs.~\eqref{WKB1} and~\eqref{S0123} into Eq.~\eqref{gKGe2} and keeping only the lowest order in $\hbar$, we get
\begin{equation}\label{HJes}
-\frac{1}{F(r)}(\partial_tS_0)^2\mathcal{P}_{t}^{2}+F(r)(\partial_rS_0)^2\mathcal{P}_{r}^{2}+\frac{1}{r^2}(\partial_{\theta}S_0)^2\mathcal{P}_{\theta}^{2}+\frac{1}{r^2\text{sin}^2\theta}(\partial_{\phi}S_0)^2\mathcal{P}_{\phi}^{2}+(\partial_{\omega}S_0)^2\mathcal{P}_{\omega}^{2}+m^2=0,
\end{equation}
where $\mathcal{P}_{\mu}$'s have been defined as
\begin{eqnarray}\label{Pmudef}
&&\mathcal{P}_{t}=1+\beta\frac{1}{F(r)}(\partial_tS_0)^2,\ \mathcal{P}_{r}=1+\beta F(r)(\partial_rS_0)^2,\ \mathcal{P}_{\theta}=1+\beta\frac{1}{r^2}(\partial_{\theta}S_0)^2, \notag\\
&&\mathcal{P}_{\phi}=1+\beta \frac{1}{r^2{\rm sin}^2\theta}(\partial_{\phi}S_0)^2,\ \mathcal{P}_{\omega}=1+\beta(\partial_{\omega}S_0)^2.
\end{eqnarray}
Eq.~\eqref{HJes} is the generalized Hamilton-Jacobi equation for scalar particles. Considering the properties of the spacetime expressed by Eq.~\eqref{metric}, we carry out separation of variables as
\begin{equation}\label{bss0fj}
 S_0 = -Et+W(r)+\Theta(\theta,\phi)+L\omega+K,
\end{equation}
where $E$ denotes the total energy of the emitted particle, $L$ represents the conserved momentum corresponding to the compact $\omega$-dimension, and $K$ is some complex constant. For the ingoing particles (moving towards the black string), the total imaginary part of $S_0$ should be zero, whereas $S_0$ of the outgoing particles (moving away from the black string) should be complex, since the outgoing trajectory is classically forbidden. Neglecting higher order terms of $\beta$ and solving Eq.~\eqref{HJes}, one obtains the solution to derivative of the radial action
\begin{equation}\label{wpr}
\partial_rW=\pm\sqrt{-\frac{m^2}{F(r)}+\frac{E^2}{F(r)^2}-\frac{J^2_{\theta}+J^2_{\phi}{\rm csc}^2\theta}{r^2 F(r)}-\frac{L^2}{F(r)}}\left(1+\frac{\mathcal{X}_1}{\mathcal{X}_2}\beta\right),
\end{equation}
where
\begin{eqnarray}
\mathcal{X}_1&=&-2F(r)m^2J^2_{\phi}+2E^2J^2_{\phi}-\frac{2F(r)J_{\theta}^2J_{\phi}^2}{r^2}-2F(r)J^2_{\phi}L^2-\frac{2F(r)J^4_{\phi}{\rm csc}^2\theta}{r^2}-F(r)m^4r^2{\rm sin}^2\theta  \notag\\
&+&2m^2r^2E^2{\rm sin}^2\theta-2F(r)m^2J_{\theta}^2{\rm sin}^2\theta+2E^2J_{\theta}^2{\rm sin}^2\theta-\frac{2F(r)J_{\theta}^4{\rm sin}^2\theta}{r^2}-2F(r)m^2r^2L^2{\rm sin}^2\theta  \notag\\
&+&2r^2E^2L^2{\rm sin}^2\theta-2F(r)J_{\theta}^2L^2{\rm sin}^2\theta-2F(r)r^2L^4{\rm sin}^2\theta,    \\
\mathcal{X}_2&=&-F(r)J_{\phi}^2-F(r)m^2r^2{\rm sin}^2\theta+r^2E^2{\rm sin}^2\theta-F(r)J_{\theta}^2{\rm sin}^2\theta-F(r)r^2L^2{\rm sin}^2\theta,
\end{eqnarray}
with $J_{\theta}=\partial_{\theta}\Theta$ and $J_{\phi}=\partial_{\phi}\Theta$.

Integrating Eq.~\eqref{wpr} around the pole at the horizon $r_{h}=2M$, we obtain the imaginary part of the radial action as
\begin{equation}\label{wrim}
Im W_\pm (r)=\pm \pi r_{h}E\left[1+2\left(m^2+\frac{J_{\theta}^2+J_{\phi}^2{\rm csc}^2\theta}{r^2_h}+L^2\right)\beta\right],
\end{equation}
where $W_{+}$ denotes the radial function of the outgoing particles and $W_{-}$ of the ingoing particles. Thus the tunneling rate of scalar particles at the event horizon is
\begin{eqnarray} \label{bstunnelproba}
\Gamma&=&\frac{P_{outgoing}}{P_{ingoing}}=\frac{{\rm exp}\left[-\frac{2}{\hbar}({\rm Im} W_{+}+{\rm Im} K)\right]}{{\rm exp}\left[-\frac{2}{\hbar}({\rm Im} W_{-}+{\rm Im} K)\right]}={\rm exp}\left[-\frac{4}{\hbar}{\rm Im} W_{+}\right] \notag\\
&=&{\rm exp}\left\{-\frac{4}{\hbar}\pi r_h E\left[1+2\left(m^2+\frac{J_{\theta}^2+J_{\phi}^2{\rm csc}^2\theta}{r^2_h}+L^2\right)\beta\right]\right\}.
\end{eqnarray}
The effective Hawking temperature for scalar particles is deduced as
\begin{equation}\label{effectiveHTs}
 T_{scalar}= \frac{\hbar}{4\pi r_h}\left[1-2\left(m^2+\frac{J_{\theta}^2+J_{\phi}^2{\rm csc}^2\theta}{r^2_h}+L^2\right)\beta\right],
\end{equation}
where $T_0 =  \frac{\hbar}{4\pi r_h}$ is the original Hawking temperature of the Schwarzschild-like black string, $\frac{J^2_{\theta}+J^2_{\phi}{\rm csc}^2\theta}{r_h^2}$ represents the kinetic energy component along the tangent plane of the horizon surface at the emission point. We can infer from Eq.~\eqref{effectiveHTs} that the corrected temperature relies on not only the mass of the black string, but also the quantum numbers
(mass, angular momentum) of the emitted scalar particles. It's obvious that the original Hawking radiation is retarded by the quantum gravity effect.

\section{Tunneling process of fermions}
\label{Section3}
We start with the Dirac Equation within HUP framework
\begin{equation}\label{dirachup}
\left(i\gamma^{\mu}\nabla_{\mu}-\frac{m}{\hbar}\right)\Psi\left(t,r,\theta,\phi,\omega\right)=0,
\end{equation}
where $\nabla_{\mu}$ is covariant derivative operator in 5-dimensional black string spacetime. Acting $i\gamma^{\nu}\nabla_{\nu}$ to Eq.~\eqref{dirachup}, one gets
\begin{equation}\label{dirachups}
\frac{1}{2}\left\{\gamma^{\mu},\gamma^{\nu}\right\}\nabla_{\mu}\nabla_{\nu}\Psi\left(t,r,\theta,\phi,\omega\right)+\frac{m^2}{\hbar^2}\Psi\left(t,r,\theta,\phi,\omega\right)=0,
\end{equation}
which should be in accordance with Klein-Gorden Equation. Then the gamma matrices should satisfy $\frac{1}{2}\left\{\gamma^{\mu},\gamma^{\nu}\right\}=-g^{\mu\nu}$, a set of 5-dimensional gamma matrices satisfy this condition shows
\begin{eqnarray}
&&\gamma^{t}=\frac{1}{\sqrt{F\left(r\right)}}\left(\begin{array}{cc}
0 & I\\
I & 0
\end{array}\right),  \gamma^{r}=\sqrt{F(r)}\left(\begin{array}{cc}
0 & \sigma^{1}\\
-\sigma^{1} & 0
\end{array}\right),  \gamma^{\theta}=\sqrt{g^{\theta\theta}}\left(\begin{array}{cc}
0 & \sigma^{2}\\
-\sigma^{2} & 0
\end{array}\right), \nonumber \\
&&\gamma^{\phi}=\sqrt{g^{\phi\phi}}\left(\begin{array}{cc}
0 & \sigma^{3}\\
-\sigma^{3} & 0
\end{array}\right), \gamma^{\omega}=i\sqrt{g^{\omega\omega}}\left(\begin{array}{cc}
I & 0\\
0 & -I
\end{array}\right),
\label{gammam}
\end{eqnarray}
where $\sigma^{i}$'s are the Pauli matrices. One can generalize Eq.~\eqref{dirachup} to MLT situation by substituting $\nabla_{\mu}$'s with $\widetilde{\nabla}_{\mu}$'s which are defined as
\begin{equation}\label{newderivative}
\widetilde{\nabla}_{0}=\nabla_{0}+\beta {\hbar}^2g^{00}\nabla_{0}^3,\ \widetilde{\nabla}_{i}=\nabla_{i}-\beta {\hbar}^2g^{ii}\nabla_{i}^3,
\end{equation}
yielding
\begin{equation}\label{diracgup}
\left(i\gamma^{\mu}\widetilde{\nabla}_{\mu}-\frac{m}{\hbar}\right)\Psi\left(t,r,\theta,\phi,\omega\right)=0.
\end{equation}

Multiplying $i\gamma^{\nu}\widetilde{\nabla}_{\nu}$ by Eq.~\eqref{diracgup}, the generalized Klein-Gorden Equation, i.e. Eq.~\eqref{gKGe2}, is recovered. It implies that fermions should share the same tunneling process with scalar particles. To further demonstrate this, we study the tunneling process of fermions starting directly from the generalized Dirac Equation, i.e. Eq.~\eqref{diracgup}. For a relativistic spin-1/2 field, there are two states corresponding respectively to spin up and spin down. Without loss of generality, we only consider the state of spin up in this paper, then the wave function of the emitted fermions can be expressed as~\cite{gaugebook}
\begin{eqnarray}
\Psi_{\uparrow}=\left(\begin{array}{c}
A\\
0\\
C\\
D
\end{array}\right)\exp\left(\frac{i}{\hbar}S\left(t,r,\theta,\phi,\omega\right)\right),
\label{wfof}
\end{eqnarray}
where $S$ is the action defined as Eq.~\eqref{S0123} with the same variables separation as~\eqref{bss0fj} and $A$, $C$ and $D$ are functions of
$t, r , \theta , \phi, \omega$. The wave function form supposed here is different with those in~\cite{{kern1,kern2,chenahep,chenjhep,chenjcap}} which suppose $D=0$, while we mention that $D=0$ only exists when the particles move along the third space direction. Inserting the wave function into the generalized Dirac equation~\eqref{diracgup}, applying the gamma matrices defined by Eq.~\eqref{gammam} and the WKB approximation, we get the the equations of motion
\begin{equation}\label{eom0}
C\frac{1}{\sqrt{F}}\partial_{t}S_0\mathcal{P}_{t}+D\sqrt{F}\partial_{r}S_0\mathcal{P}_{r}-iD\sqrt{g^{\theta\theta}}\partial_{\theta}S_0\mathcal{P}_{\theta}+C\sqrt{g^{\phi\phi}}\partial_{\phi}S_0\mathcal{P}_{\phi}+iA\sqrt{g^{\omega\omega}}\partial_{\omega}S_0\mathcal{P}_{\omega}+mA=0,
\end{equation}
\begin{equation}\label{eom1}
D\frac{1}{\sqrt{F}}\partial_{t}S_0\mathcal{P}_{t}+C\sqrt{F}\partial_{r}S_0\mathcal{P}_{r}+iC\sqrt{g^{\theta\theta}}\partial_{\theta}S_0\mathcal{P}_{\theta}-D\sqrt{g^{\phi\phi}}\partial_{\phi}S_0\mathcal{P}_{\phi}=0,
\end{equation}
\begin{equation}\label{eom2}
A\frac{1}{\sqrt{F}}\partial_{t}S_0\mathcal{P}_{t}-A\sqrt{g^{\phi\phi}}\partial_{\phi}S_0\mathcal{P}_{\phi}-iC\sqrt{g^{\omega\omega}}\partial_{\omega}S_0\mathcal{P}_{\omega}+mC=0,
\end{equation}
\begin{equation}\label{eom3}
-A\sqrt{F}\partial_{r}S_0\mathcal{P}_{r}-iA\sqrt{g^{\theta\theta}}\partial_{\theta}S_0\mathcal{P}_{\theta}-iD\sqrt{g^{\omega\omega}}\partial_{\omega}S_0\mathcal{P}_{\omega}+mD=0,
\end{equation}
where $\mathcal{P}_{\mu}$'s share the same definitions with Eq.~\eqref{Pmudef}.

One can solve Eqs.~\eqref{eom2} and~\eqref{eom3} for $C$ and $D$, obtaining
\begin{equation}\label{CDArelation}
C=\frac{\frac{1}{\sqrt{F}}\partial_{t}S_0\mathcal{P}_{t}-\sqrt{g^{\phi\phi}}\partial_{\phi}S_0\mathcal{P}_{\phi}}{i\sqrt{g^{\omega\omega}}\partial_{\omega}S_0\mathcal{P}_{\omega}-m}A,\quad D=\frac{-\sqrt{F}\partial_{r}S_0\mathcal{P}_{r}-i\sqrt{g^{\theta\theta}}\partial_{\theta}S_0\mathcal{P}_{\theta}}{i\sqrt{g^{\omega\omega}}\partial_{\omega}S_0\mathcal{P}_{\omega}-m}A;
\end{equation}
Eq.~\eqref{eom1} reflects the relation between $C$ and $D$. Inserting Eq.~\eqref{CDArelation} into Eq.~\eqref{eom0}, we get the generalized Hamilton-Jacobi equation for fermions
\begin{equation}\label{HJef}
-\frac{1}{F}(\partial_tS_0)^2\mathcal{P}_{t}^{2}+F(\partial_rS_0)^2\mathcal{P}_{r}^{2}+g^{\theta\theta}(\partial_{\theta}S_0)^2\mathcal{P}_{\theta}^{2}+g^{\phi\phi}(\partial_{\phi}S_0)^2\mathcal{P}_{\phi}^{2}+g^{\omega\omega}(\partial_{\omega}S_0)^2\mathcal{P}_{\omega}^{2}+m^2=0,
\end{equation}
which is consistent with Eq.~\eqref{HJes}. The calculation for the spin down state is parallel, one can repeat the process to get the same result as spin up case.

Along the same path as in section ~\ref{Section2}, the effective Hawking temperature for fermions should be equal to that for scalar particles
\begin{equation}\label{effectiveHTf}
 T_{fermions}= \frac{\hbar}{4\pi r_h}\left[1-2\left(m^2+\frac{J_{\theta}^2+J_{\phi}^2{\rm csc}^2\theta}{r^2_h}+L^2\right)\beta\right].
\end{equation}

\section{Tunneling process of massive bosons}
\label{Section4}

The MLT-corrected equation of motion of massive bosons was derived in~\cite{Li:2016mwq}. For the case of uncharged bosons or uncharged spacetime background, it yields
\begin{equation} \label{feomb}
\partial_{\mu}\left(\sqrt{-g}\mathfrak{B}^{\mu\nu}\right)-\sqrt{-g}\frac{m^2}{\hbar^2}\mathfrak{B}^{\nu}+\beta\hbar^2\partial_{0}\partial_{0}\partial_{0}\left(\sqrt{-g}g^{00}\mathfrak{B}^{0\nu}\right)-\beta\hbar^2\partial_{i}\partial_{i}\partial_{i}\left(\sqrt{-g}g^{ii}\mathfrak{B}^{i\nu}\right)=0.
\end{equation}

According to the WKB approximation, the boson field $\mathfrak{B}_\mu$ is of the form
\begin{equation} \label{WKBmb}
\mathfrak{B}_\mu=C_\mu(t, r, \theta, \phi, \omega) {\rm exp}\left[\frac{i}{\hbar}S(t, r, \theta, \phi, \omega)\right],
\end{equation}
where $S$ is also defined as Eq.~\eqref{S0123}. Substituting Eqs.~\eqref{WKBmb},~\eqref{S0123} into Eq.~\eqref{feomb}, and keeping only the lowest order in $\hbar$, we get equations of the coefficients $C_\mu$
\begin{align}\label{HJeomb0}
&C_0 m^2+ g^{rr}[C_0 (\partial_rS_0)^2\mathcal{P}_{r}^{2}-C_1(\partial_tS_0)(\partial_rS_0)\mathcal{P}_{t}\mathcal{P}_{r}]+g^{\theta\theta}[C_0 (\partial_{\theta}S_0)^2\mathcal{P}_{\theta}^{2}-C_2(\partial_tS_0)(\partial_{\theta}S_0)\mathcal{P}_{t}\mathcal{P}_{\theta}] \nonumber\\
&+g^{\phi\phi}[C_0 (\partial_{\phi}S_0)^2\mathcal{P}_{\phi}^{2}-C_3(\partial_tS_0)(\partial_{\phi}S_0)\mathcal{P}_{t}\mathcal{P}_{\phi}]+g^{\omega\omega}[C_0 (\partial_{\omega}S_0)^2\mathcal{P}_{\omega}^{2}-C_4(\partial_tS_0)(\partial_{\omega}S_0)\mathcal{P}_{t}\mathcal{P}_{\omega}]=0,
\end{align}
\begin{align}\label{HJeomb1}
&C_1 m^2+ g^{tt}[C_1 (\partial_tS_0)^2\mathcal{P}_{t}^{2}-C_0(\partial_rS_0)(\partial_tS_0)\mathcal{P}_{r}\mathcal{P}_{t}]+g^{\theta\theta}[C_1 (\partial_{\theta}S_0)^2\mathcal{P}_{\theta}^{2}-C_2(\partial_rS_0)(\partial_{\theta}S_0)\mathcal{P}_{r}\mathcal{P}_{\theta}]  \nonumber\\
&+g^{\phi\phi}[C_1 (\partial_{\phi}S_0)^2\mathcal{P}_{\phi}^{2}-C_3(\partial_rS_0)(\partial_{\phi}S_0)\mathcal{P}_{r}\mathcal{P}_{\phi}]+g^{\omega\omega}[C_1 (\partial_{\omega}S_0)^2\mathcal{P}_{\omega}^{2}-C_4(\partial_rS_0)(\partial_{\omega}S_0)\mathcal{P}_{r}\mathcal{P}_{\omega}]=0,
\end{align}
\begin{align}\label{HJeomb2}
&C_2 m^2+ g^{tt}[C_2 (\partial_tS_0)^2\mathcal{P}_{t}^{2}-C_0(\partial_{\theta}S_0)(\partial_tS_0)\mathcal{P}_{\theta}\mathcal{P}_{t}]+g^{rr}[C_2 (\partial_{r}S_0)^2\mathcal{P}_{r}^{2}-C_1(\partial_{\theta}S_0)(\partial_{r}S_0)\mathcal{P}_{\theta}\mathcal{P}_{r}]  \nonumber\\
&+g^{\phi\phi}[C_2 (\partial_{\phi}S_0)^2\mathcal{P}_{\phi}^{2}-C_3(\partial_{\theta} S_0)(\partial_{\phi}S_0)\mathcal{P}_{\theta}\mathcal{P}_{\phi}]+g^{\omega\omega}[C_2 (\partial_{\omega}S_0)^2\mathcal{P}_{\omega}^{2}-C_4(\partial_{\theta}S_0)(\partial_{\omega}S_0)\mathcal{P}_{\theta}\mathcal{P}_{\omega}]=0,
\end{align}
\begin{align}\label{HJeomb3}
&C_3 m^2+ g^{tt}[C_3 (\partial_tS_0)^2\mathcal{P}_{t}^{2}-C_0(\partial_{\phi}S_0)(\partial_tS_0)\mathcal{P}_{\phi}\mathcal{P}_{t}]+g^{rr}[C_3 (\partial_{r}S_0)^2\mathcal{P}_{r}^{2}-C_1(\partial_{\phi}S_0)(\partial_{r}S_0)\mathcal{P}_{\phi}\mathcal{P}_{r}]  \nonumber\\
&+g^{\theta\theta}[C_3 (\partial_{\theta}S_0)^2\mathcal{P}_{\theta}^{2}-C_2(\partial_{\phi} S_0)(\partial_{\theta}S_0)\mathcal{P}_{\phi}\mathcal{P}_{\theta}]+g^{\omega\omega}[C_3 (\partial_{\omega}S_0)^2\mathcal{P}_{\omega}^{2}-C_4(\partial_{\phi}S_0)(\partial_{\omega}S_0)\mathcal{P}_{\phi}\mathcal{P}_{\omega}]=0,
\end{align}
\begin{align}\label{HJeomb4}
&C_4 m^2+ g^{tt}[C_4 (\partial_tS_0)^2\mathcal{P}_{t}^{2}-C_0(\partial_{\omega}S_0)(\partial_tS_0)\mathcal{P}_{\omega}\mathcal{P}_{t}]+g^{rr}[C_4 (\partial_{r}S_0)^2\mathcal{P}_{r}^{2}-C_1(\partial_{\omega}S_0)(\partial_{r}S_0)\mathcal{P}_{\omega}\mathcal{P}_{r}]  \nonumber\\
&+g^{\theta\theta}[C_4 (\partial_{\theta}S_0)^2\mathcal{P}_{\theta}^{2}-C_2(\partial_{\omega} S_0)(\partial_{\theta}S_0)\mathcal{P}_{\omega}\mathcal{P}_{\theta}]+g^{\phi\phi}[C_4 (\partial_{\phi}S_0)^2\mathcal{P}_{\phi}^{2}-C_3(\partial_{\omega}S_0)(\partial_{\phi}S_0)\mathcal{P}_{\omega}\mathcal{P}_{\phi}]=0,
\end{align}
where $\mathcal{P}_{\mu}$'s definition in Eq.~\eqref{Pmudef} has been employed again.  Applying the variables separation~\eqref{bss0fj} and the black string metric~\eqref{metric}, we reformulate Eqs.~\eqref{HJeomb0}-\eqref{HJeomb4} to a matrix equation $K\left(C_{0},C_{1},C_{2},C_{3},C_{4}\right)^T=0$ and the elements of $5\times5$ matrix $K$ are expressed as
\begin{eqnarray}
&&K_{11}= F(r)W'^2\mathcal{P}_{r}^{2}+\frac{{J_{\theta}}^2\mathcal{P}_{\theta}^{2}}{r^2}+\frac{{J_{\phi}}^2\mathcal{P}_{\phi}^{2}}{r^2{\rm sin}^2\theta}+L^2\mathcal{P}_{\omega}^{2}+m^2,\ K_{12}=F(r)EW'\mathcal{P}_{t}\mathcal{P}_{r},\notag\\
&&K_{13}=\frac{EJ_{\theta}\mathcal{P}_{t}\mathcal{P}_{\theta}}{r^2},\ K_{14}=\frac{EJ_{\phi}\mathcal{P}_{t}\mathcal{P}_{\phi}}{r^2{\rm sin}^2\theta},\ K_{15}=EL\mathcal{P}_{t}\mathcal{P}_{\omega}, \ K_{21}=-\frac{W'E\mathcal{P}_{r}\mathcal{P}_{t}}{F(r)},\notag \\
&&K_{22}=-\frac{E^2\mathcal{P}_{t}^{2}}{F(r)}+\frac{{J_{\theta}}^2\mathcal{P}_{\theta}^{2}}{r^2}+\frac{{J_{\phi}}^2\mathcal{P}_{\phi}^{2}}{r^2{\rm sin}^2\theta}+L^2\mathcal{P}_{\omega}^{2}+m^2,\ K_{23}=\frac{W'J_{\theta}\mathcal{P}_{r}\mathcal{P}_{\theta}}{r^2},\notag\\
&&K_{24}=\frac{W'J_{\phi}\mathcal{P}_{r}\mathcal{P}_{\phi}}{r^2{\rm sin}^2\theta},\ K_{25}=W'L\mathcal{P}_{r}\mathcal{P}_{\omega},\ K_{31}=-\frac{J_{\theta}E\mathcal{P}_{\theta}\mathcal{P}_{t}}{F(r)},\ K_{32}=F(r)J_{\theta}W'\mathcal{P}_{\theta}\mathcal{P}_{r}, \notag \\
&&K_{33}=-\frac{E^2\mathcal{P}_{t}^{2}}{F(r)}+F(r)W'^2\mathcal{P}_{r}^{2}+\frac{{J_{\phi}}^2\mathcal{P}_{\phi}^{2}}{r^2{\rm sin}^2\theta}+L^2\mathcal{P}_{\omega}^{2}+m^2,\ K_{34}=\frac{J_{\theta}J_{\phi}\mathcal{P}_{\theta}\mathcal{P}_{\phi}}{r^2{\rm sin}^2\theta}, \notag\\
&&K_{35}=J_{\theta}L\mathcal{P}_{\theta}\mathcal{P}_{\omega},\ K_{41}=-\frac{J_{\phi}E\mathcal{P}_{\phi}\mathcal{P}_{t}}{F(r)},\ K_{42}=F(r)J_{\phi}W'\mathcal{P}_{\phi}\mathcal{P}_{r},\ K_{43}=\frac{J_{\phi}J_{\theta}\mathcal{P}_{\phi}\mathcal{P}_{\theta}}{r^2}, \notag\\
&&K_{44}=-\frac{E^2\mathcal{P}_{t}^{2}}{F(r)}+F(r)W'^2\mathcal{P}_{r}^{2}+\frac{{J_{\theta}}^2\mathcal{P}_{\theta}^{2}}{r^2}+L^2\mathcal{P}_{\omega}^{2}+m^2,\ K_{45}=J_{\phi}L\mathcal{P}_{\phi}\mathcal{P}_{\omega}, \notag \\
&&K_{51}=-\frac{LE\mathcal{P}_{\omega}\mathcal{P}_{t}}{F(r)},\ K_{52}=F(r)LW'\mathcal{P}_{\omega}\mathcal{P}_{r},\ K_{53}=\frac{LJ_{\theta}\mathcal{P}_{\omega}\mathcal{P}_{\theta}}{r^2},\ K_{54}=\frac{LJ_{\phi}\mathcal{P}_{\omega}\mathcal{P}_{\phi}}{r^2{\rm sin}^2\theta}, \notag \\
&&K_{55}=-\frac{E^2\mathcal{P}_{t}^{2}}{F(r)}+F(r)W'^2\mathcal{P}_{r}^{2}+\frac{{J_{\theta}}^2\mathcal{P}_{\theta}^{2}}{r^2}+\frac{{J_{\phi}}^2\mathcal{P}_{\phi}^{2}}{r^2{\rm sin}^2\theta}+m^2,
\end{eqnarray}
where $\partial_{\mu}S_0$'s in $\mathcal{P}_{\mu}$'s should be also replaced by $(-E, W', J_{\theta}, J_{\phi}, L)$.

The determination of the coefficient matrix should be equal to zero to ensure that the matrix equation possesses nontrivial solution. It yields the following equation
\begin{eqnarray}\label{bsWequation}
&&8F(r)^5m^2\beta(\partial_rW)^{10}+\left[24F(r)^4m^2\left(m^2-\frac{E^2}{F(r)}+\frac{J^2_{\theta}+J^2_{\phi}{\rm csc}^2\theta}{r^2}+L^2\right)\beta+F(r)^4m^2\right](\partial_rW)^{8}\notag\\
&&+(A_{6}\beta+B_{6})(\partial_rW)^{6}+(A_{4}\beta+B_{4})(\partial_rW)^{4}+(A_{2}\beta+B_{2})(\partial_rW)^{2}+A_{0}\beta+B_0\simeq0,
\end{eqnarray}
where the terms of higher order of $\beta$ have been omitted, the unspecified parameters $A_{i}$'s and $B_{i}$'s are both complicated function of $m, E, J_{\theta}, J_{\phi}, L$ and the black string metric, thus we don't write them down here. Solving Eq.~\eqref{bsWequation}, we get the solution to derivative of the radial action
\begin{equation}\label{wprmb}
\partial_rW=\pm\sqrt{-\frac{m^2}{F(r)}+\frac{E^2}{F(r)^2}-\frac{J^2_{\theta}+J^2_{\phi}{\rm csc}^2\theta}{r^2 F(r)}-\frac{L^2}{F(r)}}\left(1+\frac{\mathcal{Y}_1}{\mathcal{Y}_2}\beta\right),
\end{equation}
where
\begin{eqnarray}
\mathcal{Y}_1&=&-8F(r)m^2J^2_{\phi}+8E^2J^2_{\phi}-\frac{9F(r)J^2_{\theta}J^2_{\phi}}{r^2}-\frac{3F(r)J^4_{\theta}J^2_{\phi}}{2m^2r^4}-8F(r)J^2_{\phi}L^2-\frac{7F(r)J^4_{\phi}{\rm csc}^2\theta}{r^2} \notag\\
&+&\frac{3F(r)J^2_{\theta}J^4_{\phi}{\rm csc}^2\theta}{2m^2r^4}-4F(r)m^4r^2{\rm sin}^2\theta+8m^2r^2E^2{\rm sin}^2\theta-8F(r)m^2J^2_{\theta}{\rm sin}^2\theta+8E^2J^2_{\theta}{\rm sin}^2\theta  \notag\\
&-&\frac{8F(r)J^4_{\theta}{\rm sin}^2\theta}{r^2}-8F(r)m^2r^2L^2{\rm sin}^2\theta+8r^2E^2L^2{\rm sin}^2\theta-8F(r)J^2_{\theta}L^2{\rm sin}^2\theta \notag\\
&-&4F(r)r^2J^2_{\phi}L^2{\rm sin}^2\theta-4F(r)r^2L^2{\rm sin}^2\theta,  \\
\mathcal{Y}_2&=&-F(r)J_{\phi}^2-F(r)m^2r^2{\rm sin}^2\theta+r^2E^2{\rm sin}^2\theta-F(r)J_{\theta}^2{\rm sin}^2\theta-F(r)r^2L^2{\rm sin}^2\theta.
\end{eqnarray}

Integrating Eq.~\eqref{wprmb} around the pole at the horizon $r_{h}=2M$, we obtain the imaginary part of the radial action for massive bosons as
\begin{equation}\label{wrim}
Im W_\pm (r)=\pm \pi r_{h}E\left[1+8\left(m^2+\frac{J_{\theta}^2+J_{\phi}^2{\rm csc}^2\theta}{r^2_h}+L^2\right)\beta\right],
\end{equation}
where $W_{+}$ denotes the radial function of the outgoing particles and $W_{-}$ of the ingoing particles. Thus the tunneling rate of massive bosons at the event horizon is
\begin{eqnarray} \label{bstunnelproba}
\Gamma&=&\frac{P_{outgoing}}{P_{ingoing}}=\frac{{\rm exp}\left[-\frac{2}{\hbar}({\rm Im} W_{+}+{\rm Im} K)\right]}{{\rm exp}\left[-\frac{2}{\hbar}({\rm Im} W_{-}+{\rm Im} K)\right]}={\rm exp}\left[-\frac{4}{\hbar}{\rm Im} W_{+}\right] \notag\\
&=&{\rm exp}\left\{-\frac{4}{\hbar}\pi r_h E\left[1+8\left(m^2+\frac{J_{\theta}^2+J_{\phi}^2{\rm csc}^2\theta}{r^2_h}+L^2\right)\beta\right]\right\}.
\end{eqnarray}
The effective Hawking temperature for massive bosons is deduced as
\begin{equation}\label{effectiveHTb}
 T_{bosons}= \frac{\hbar}{4\pi r_h}\left[1-8\left(m^2+\frac{J_{\theta}^2+J_{\phi}^2{\rm csc}^2\theta}{r^2_h}+L^2\right)\beta\right].
\end{equation}
Comparing Eq.~\eqref{effectiveHTb} to Eqs.~\eqref{effectiveHTs} and~\eqref{effectiveHTf}, we find that the temperature correction for massive bosons is four times as big as that for scalar particles and fermions, given the same particle mass and angular momentum components.

\section{Thermodynamics of black string}
\label{Section5}
\begin{figure}[!ht]
\centering
\includegraphics[scale=1.5]{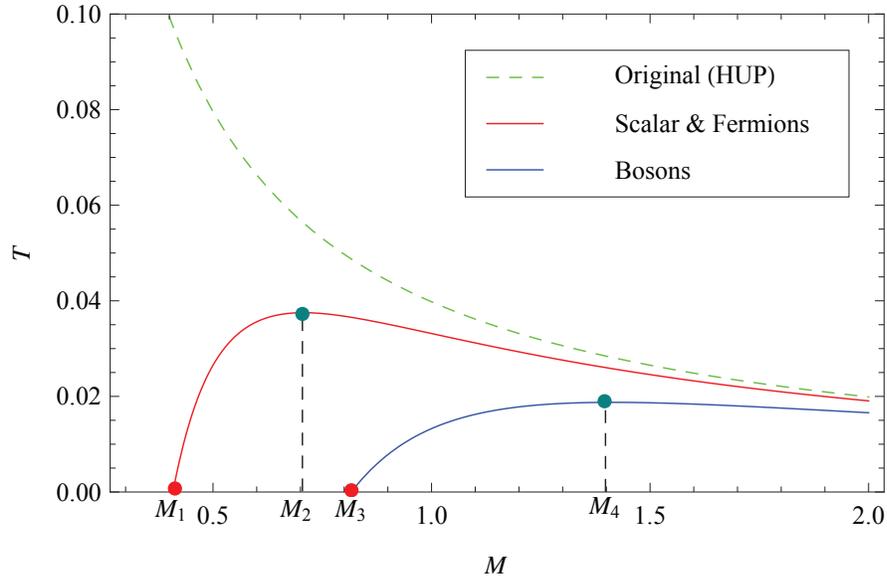}
\caption{Temperature of black string versus its mass. Mass is in units of $m^2_P/M_f$ and temperature is in units of $c^2 M_f /k_B$. The green (dashed) curve is the original Hawking temperature in HUP framework; The rad (upper) solid curve is the MLT-corrected effective Hawking temperature for scalar particles and fermions, and the blue (lower) solid one is for bosons.}
\label{fig:1}       
\end{figure}

In this section, we study the thermodynamics of the Schwarzschild-like black string with the discussions above. For this, we rewrite Eqs.~\eqref{effectiveHTs},~\eqref{effectiveHTf} and~\eqref{effectiveHTb} as
\begin{equation}\label{effectiveHT1}
T=\frac{c^2m^2_P}{8\pi k_BM}\left(1-\frac{aE^2}{3c^4M^2_f}\right)
\end{equation}
by approximating $m^2+\frac{J^2_{\theta}+J^2_{\phi}{\rm csc}^2\theta}{r_h^2}+L^2$ as $E^2$, where $a=2$ for scalar particles and fermions, while $a=8$ for bosons. The particles near the black string surface have an intrinsic position uncertainty of about the horizon radius $\frac{2GM}{c^2}$~\cite{adlerGRG}, then we have the relations
\begin{equation}\label{Erelations}
E\sim c\Delta p\sim c\frac{\hbar}{\triangle x}\sim c\frac{\hbar}{r_h},
\end{equation}
where the standard uncertainty principle has been used. Replacing $E$ with $\frac{c\hbar}{r_h}$ in Eq.~\eqref{effectiveHT1}, one obtains
\begin{equation}\label{effectiveHT2}
T=\frac{c^2m^2_P}{8\pi k_BM}\left(1-\frac{am^4_P}{12M^2_fM^2}\right).
\end{equation}
The behavior of the MLT-corrected temperature and the original Hawking temperature are plotted in Figure~\ref{fig:1}. Both temperature curves for scalar particles/fermions and massive bosons are bulge-shaped. The MLT-corrected temperature for bosons reduces to zero at $M_3=\sqrt{\frac{2}{3}}\frac{m^2_P}{M_f}$, and that for scalar particles and fermions shrinks to zero at $M_1=\sqrt{\frac{1}{6}}\frac{m^2_P}{M_f}$, well before the black string evaporate completely.

The corresponding black string entropy can be calculated as
\begin{eqnarray}\label{entropy}
S &=& \int\frac{c^2dM}{k_BT} \nonumber\\
  &=& \frac{4\pi}{m^2_P}M^2+\frac{2\pi am^2_P}{3M^2_f}{\rm ln}M  \nonumber\\
  &=& \frac{c^4A}{4G^2m^2_P}+\frac{\pi am^2_P}{3M^2_f}{\rm ln}\frac{c^4A}{16\pi G^2},
\end{eqnarray}
where $A=4\pi r^2_h$ is the area of the horizon surface. Note that the MLT effect contribute a positive ${\rm ln}A$ correction to the entropy, different than a negative ${\rm ln}A$ correction suggested in~\cite{majumPLB,binaPRD,chenNPPS,adlerGRG,xiangwenJHEP,majumGRG}.

\begin{figure}[!ht]
\centering
\includegraphics[scale=1.5]{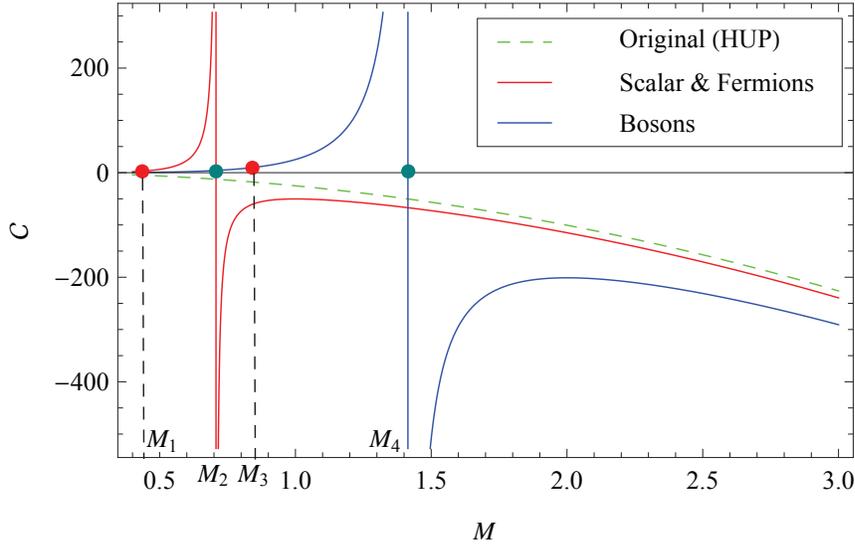}
\caption{Heat capacity of black string versus its mass. Mass is in units of $m^2_P/M_f$ and heat capacity is in units of $(m_P/M_f)^2 k_B$. The green (dashed) curve is the original heat capacity in HUP framework; The rad (left) solid curve is the MLT-corrected heat capacity for scalar particles and fermions, and the blue (right) solid one is for bosons.}
\label{fig:2}       
\end{figure}

The MLT-corrected heat capacity of the black string is given by
\begin{eqnarray}\label{heatcapacity}
\mathcal{C} &=& c^2\frac{dM}{dT} \nonumber\\
  &=& \frac{8\pi k_B}{-\frac{m^2_P}{M^2}+\frac{a m^6_P}{4M^2_fM^4}},
\end{eqnarray}
which is plotted in Figure.~\ref{fig:2}. We observe that heat capacities for scalar particles/fermions and bosons diverge respectively at $M_2=\sqrt{\frac{1}{2}}\frac{m^2_P}{M_f}$ and $M_4=\sqrt{2}\frac{m^2_P}{M_f}$, where the MLT-corrected temperatures take maximums (dark green points in Figure.~\ref{fig:1}). The heat capacities are positive at $M_1$ and $M_3$, where the MLT-corrected temperatures reduce to zero.

Then we can represent the evaporation process of the Schwarzschild-like black string qualitatively. When $M>M_4$, both the corrected temperatures increase with negative heat capacity as the black string mass decrease, meaning increasing tunneling rates of the particle crossing the horizon; The tunneling rate of bosons reaches its maximum at $M=M_4$ when the corrected temperature for bosons maximizes; When $M_4>M>M_3$, the corrected temperature for bosons decrease; The black string cease to emit bosons at $M=M_3$; When $M<M_3$, only scalar particles and fermions can tunnel from the black string, their corrected temperatures maximize at $M=M_2$ and then decrease with positive heat capacity after that; Eventually, both the scalar particles and fermions cease being emitted from the black string at $M=M_1$. Hereafter, the black string never evaporates any more. Once the black string could evaporate particles again, its temperature increases, then the mass of black string should also increase because of the positive heat capacities for all kinds of particles in the vicinity of $M_1$, which is in contradiction to the fact that evaporation should decrease the mass of black string. It seems that the existence of minimal length and time-interval should lead to a black string remnant.

However we should be careful when we predict the final state of the black string evaporation. As the black string mass reduces, its size, characterized by two times horizon radius, could reach the minimal length allowed in LXDs model $L_f$ , which is related to $M_f$ by $L_f M_f=\hbar$. We stress that the tunneling method adopted in this work to study the evaporation process only makes sense when
\begin{equation}\label{minmassGUP}
\frac{4GM}{c^2}>\frac{m_P}{M_f}\sqrt{\frac{\hbar G}{c^3}},
\end{equation}
yielding $M>M_{cri.}=\frac{1}{4}\frac{m^2_P}{M_f}$. Once the black string mass reduces to $M_{cri.}$, the classical or semi-classical concept of black string should be no longer valid. Since the black string horizon is ill-defined, one can never tell whether or not a particle is ``in" the black string, thus the tunneling method becomes inefficient to study black string evaporation. Note that $M_1$ and $M_3$ have the same order of magnitude with $M_{cri.}$, we conclude that a full quantum gravity theory is needed to study the last stage of black string evaporation and verify the outcomes we obtained in this section. Recently, an interesting quantum gravity theory with spin and scaling gauge symmetries has been proposed by Wu~\cite{Wu:2015wwa,Wu:2015hoa}.

\section{Discussion and conclusion}
\label{Section6}
In this paper, incorporating the minimal length and time-interval effect, we have studied the tunneling process of scalar particles, fermions and massive bosons from the 5-dimensional Schwarzschild-like black string, respectively. The original Hawking radiation was retarded by the quantum gravity effect for all these kinds of particles. The generalized Hamilton-Jacobi equations for the scalar particles and fermions were showed to be the same by processing the generalized Klein-Gorden and Dirac Equations, thus they shared the same effective Hawking temperatures. The temperature correction for bosons was four times as big as that for scalar particles and fermions. By observing the effective Hawking temperatures and heat capacities for all kinds of particles, the evaporation process of the Schwarzschild-like black string was represented qualitatively. The bosons ceased tunneling from the black string at first, then the scalar particles and fermions ceased being radiated as the mass of black string decreased to $\sqrt{\frac{1}{6}}\frac{m^2_P}{M_f}$. Even a black string remnant were expected as the consequence of evaporation in our analysis, we pointed that the final stage of black string evaporation should be reexamined by full quantum gravity theory.

\appendix
\section{Feasibility analysis of applying WKB approximation to studying the quantum gravity effects on particles tunneling process}
\label{appendix1}
As pointed out in the introduction part, WKB approximation retains only the leading order of $\hbar$ in the classical action of the trajectory. However, quantum gravity effect which matters in Planck scale should be difficult to identify in low energy limit. It should be ensured that WKB approximation is precise enough to let the particles ``feel" the effect of quantum gravity when we study the black string/black hole evaporation. Next, we analyze the scalar particles case as an example.

Inserting Eq.~\eqref{WKB1} into Eq.~\eqref{gKGe2}, and keeping the terms in $S$ to the first order of $\hbar$, i.e. $S=S_0+\hbar S_1$, one obtains
\begin{eqnarray}\label{gKGe3}
&&g^{\mu\nu}[\partial_{\mu}S_0\partial_{\nu}S_0+\hbar(\partial_{\mu}S_0\partial_{\nu}S_1+\partial_{\nu}S_0\partial_{\mu}S_1-i\partial_{\mu}\partial_{\nu}S_0)+(-1)^{\delta^0_{\nu}}\beta\partial_{\mu}S_0g^{\nu\nu}(\partial_{\nu}S_0)^3 \nonumber \\
&&+(-1)^{\delta^0_{\mu}}\beta\partial_{\nu}S_0g^{\mu\mu}(\partial_{\mu}S_0)^3+\mathcal{O}(\beta\hbar)+\mathcal{O}(\hbar^2)+\mathcal{O}(\beta^2)+\cdots]+m^2=0.
\end{eqnarray}
We observe that the leading order terms of $\hbar$ at the left hand side of Eq.~\eqref{gKGe3} are of the forms $g^{\mu\nu}\partial_{\mu}S_0\partial_{\nu}S_1\hbar$ and $ig^{\mu\nu}\partial_{\mu}\partial_{\nu}S_0\hbar$. These terms should be much smaller than the leading order terms of $\beta$, which are of the form $g^{\mu\nu}\partial_{\mu}S_0\partial_{\nu}S_0g^{\nu\nu}(\partial_{\nu}S_0)^2\beta$. We suppose $\partial_{\mu}S_0\sim\partial_{\mu}S_1\sim\partial_{\mu}\partial_{\nu}S_0$ at the level of numerical value. For $\mu=\nu=0$, we have the relation
\begin{equation}\label{restrict}
\frac{E^2}{3M^2_fc^4}\gg\frac{\hbar}{1{\rm J\cdot s}}, \quad \frac{E^3}{3M^2_fc^4}\cdot\frac{1}{1{\rm J}}\gg\frac{\hbar}{1{\rm J\cdot s}},
\end{equation}
which yield $E\gg10^{-17}{\rm Tev}$ and $E\gg10^{-9}{\rm Tev}$ in turn, given $M_f\sim1{\rm Tev}$. Relates $E$ with the black string mass by $E\sim\frac{\hbar c^3}{2GM}$, as discussed in section~\ref{Section5}, then the black string mass should satisfy $M\ll 10^{17}\frac{m^2_P}{M_f}$ and $M\ll 10^{9}\frac{m^2_P}{M_f}$, respectively. All these conditions for the energy of the emitted particles $E$ and the black string mass $M$ are well satisfied, thus it's sufficient to retain only the leading order term of $\hbar$ in the classical action of trajectory when studying black string/black hole evaporation. The application of WKB approximation in this work is justified.

\section*{Acknowledgments}
This work was supported by the National Science Foundation of China (NSFC) under Grant No. 11475237, No. 11121064 and No. 10821504.

\end{document}